\documentclass{article}
\usepackage{setspace}
\usepackage{anysize}
\marginsize{3cm}{3cm}{3cm}{5cm}

\begin{document}

\date{} 

\begin{doublespace} 

\title{Relaxation time for the temperature in a dilute binary mixture from classical kinetic theory}
\maketitle

\begin{center}
\author{Valdemar Moratto $^{a)}$, L.S. Garc\'ia-Col\'in $^{a),b)}$}\\

\emph{$^{a)}$ Departamento de F\'isica, Universidad Aut\'onoma Metropolitana-Iztapalapa, Av. San Rafael Atlixco 186, Col Vicentina, 09340, M\'exico D. F. M\'exico.}\\

\emph{$^{b)}$ El Colegio Nacional, M\'exico D. F., M\'exico.}
\end{center}
(Received
\begin{abstract}
The system of our interest is a dilute binary mixture, in which we consider that the species have different temperatures as an initial condition. To study their time evolution, we use the full version of the Boltzmann equation, under the hypothesis of partial local equilibrium for both species. Neither a diffusion force nor mass diffusion appears in the system. We also estimate
the time in which the temperatures of the components reach the full local equilibrium.
In solving the Boltzmann equation, we imposed no assumptions on the collision term. We work out its
solution by using the well known Chapman-Enskog method to first order in the gradients. The time in
which the temperatures relax is obtained following Landau's original idea. The result is that the relaxation time for the temperatures is much smaller than the characteristic hydrodynamical times but greater than a collisional time. The main conclusion is that there is no need to study binary mixtures with different temperatures when hydrodynamical properties are sought.
\end{abstract}


\newpage
\section{Introduction}
Binary mixtures in which the temperatures are equal have been treated in literature for dilute gases \cite{Chapman-Cowling,teoria cinetica LGCS,Fisica de los Procesos Irreversibles}, both in the case of inert components as well as those composed by charged particles thus leading to the study of plasmas \cite{Spitzer,Braginski,Balescu I,Balescu II}. These have been studied with kinetic theory when temperatures of ions and electrons are equal \cite{Marshall,Plasmas Grupo I,Plasmas Grupo II,Libro Plasmas}, and other authors.

The question about the time relaxation for a binary mixture at different temperatures has been treated by Landau \cite{Landau,Physical Kinetics}. He used a Fokker-Planck kinetic equation. In this work we use the complete Boltzmann equation with no assumptions on the collisional term. We use the Chapman-Enskog method instead Grad's method because it is better defined in physical terms \cite{Grad,inconsistencia Grad}. This allow us to give a microscopic foundation about the relaxation time. Such a calculation can be extended to plasmas but in discussion section we will give a few arguments that makes the exercise unnecessary.
\\

When we solve the Boltzmann equation with the Chapman-Enskog expansion to first order in gradients \cite{Chapman-Cowling} there is no diffusion force appearing in the linearized Boltzmann equation. This result deserves more discussion because it has generated misunderstandings in the literature \cite{chinos}.
\\

This paper is divided as follows: Section 2 is devoted to some definitions and a review of kinetic theory
as it relates to the problem. In section 3 we discuss the solution method and exhibit the linearized Boltzmann equation. In section 4 the time in which the temperatures reach the full local equilibrium is estimated.
Section 5 is devoted to some concluding remarks.

\section{The kinetic theory scenario}
We consider a dilute binary mixture of particles with masses $m_{a}$ and $m_{b}$, The density is $n=n_{a}+n_{b}$, so the mass density is $\rho=\rho_a+\rho_b=m_{a}n_{a}+m_{b}n_{b}$. Each species has a different local temperature, $T_{i}$ where $i=a,b$. The Boltzmann equation in the absence of external forces for the evolution of the distribution function for each species is
\begin{eqnarray}
\frac{\partial f_{i}}{\partial t} +\textbf{v}_{i}\cdot\frac{\partial f_{i}}{\partial\textbf{r}} =\sum_{i,j=a}^{b}J(f_{i}f_{j})\label{BE}
\end{eqnarray}
where
\begin{eqnarray}
J(f_{i}f_{j}) =\int\cdots\int \left[f(\textbf{v}_{i}')f(\textbf{v}_{j}') -f(\textbf{v}_{i})f(\textbf{v}_{j})\right]\label{colisionalkernel}\\\nonumber
\times \sigma\left(\textbf{v}_{i}\textbf{v}_{j}\rightarrow \textbf{v}_{i}'\textbf{v}_{j}'\right) g_{ij} d\textbf{v}_{j} d \textbf{v}_{i}' d\textbf{v}_{j}',
\end{eqnarray}
where $g_{ij}$ is the relative velocity between two particles when they collide. We remind the reader that there is a Boltzmann equation with the same structure for each species, coupled by $J(f_if_j)$. The cross section $\sigma$ satisfies the principle of microscopic reversibility guaranteeing the existence of inverse collisions.
\\

Next we introduce some useful definitions. In general the average of any dynamical variable $\psi_{i}$ is given by
\begin{eqnarray}
\langle\psi_{i}\rangle\equiv\psi_{i}(\textbf{r},t) =\frac{1}{n_{i}}\int\psi_{i}(\textbf{r},\textbf{v}_{i},t) f_{i}(\textbf{r},\textbf{v}_{i},t) d\textbf{v}_{i},
\end{eqnarray}
as usual we have the local particle densities
\begin{eqnarray}
n_{i}(\textbf{r},t)=\int f_{i}(\textbf{r},\textbf{v}_{i},t) d\textbf{v}_{i}.
\end{eqnarray}
The local barycentric velocity is given by
\begin{eqnarray}
\rho\textbf{u}(\textbf{r},t) =\sum_{i}\rho_{i}\textbf{u}_{i}(\textbf{r},t),
\end{eqnarray}
where the local velocity of each species is
\begin{eqnarray}
\textbf{u}_{i}(\textbf{r},t)=\frac{1}{n_{i}}\int f_{i}(\textbf{r},\textbf{v}_{i},t) \textbf{v}_{i} d\textbf{v}_{i},
\end{eqnarray}
implying that the chaotic velocity is defined as $\textbf{c}_{i}\equiv\textbf{v}_{i}-\textbf{u}_{i}$. Recall that here we are taking $\textbf{u}_{i}$ as a local variable, because our hypothesis is that each species is in local equilibrium by itself. This is rather crucial in the calculation of the transport properties of the mixture since the mass flux $\textbf{J}_{i}=\overrightarrow{0}$ for $i=a,b$. The barycentric velocity $\textbf{u}$ is not a candidate to be chosen as a local variable, because the temperatures of each species is different.
\\

The kinetic part of the stress tensor for the species $i$ is defined as
\begin{eqnarray}
\tau_{i}= m_{i}\int f_{i}\textbf{c}_{i}\textbf{c}_{i} d\textbf{v}_{i}.
\label{tensor esfuerzos}
\end{eqnarray}
Also, the heat flux for each species is given by:
\begin{eqnarray}
\textbf{J}_{q_{i}}=\frac{1}{2}\rho_{i}\langle \textbf{c}_{i}c_{i}^{2}\rangle,
\label{heat flux}
\end{eqnarray}
and the mass flux is given by
\begin{eqnarray}
\textbf{J}_{i}=\rho_{i}\langle\textbf{c}_{i}\rangle=0
\end{eqnarray}
meaning that there is no self-diffusion because the local velocity that we have taken is not the barycentric velocity of the mixture. Here $\textbf{c}_{i}$ is the so called thermal or chaotic velocity. The average internal energy for the species $i$ is,
\begin{eqnarray}
\rho_{i}\varepsilon_{i}=\frac{1}{2}\rho_{i}\langle c_{i}^{2}\rangle,
\end{eqnarray}
which allows the introduction of the corresponding kinetic temperatures, namely,
\begin{eqnarray}
\varepsilon_{i}=\frac{3}{2}n_{i}kT_{i}.
\end{eqnarray}
Recall that the averages that represent macroscopic quantities are performed over $\textbf{c}_i$ and not over $\textbf{v}_i$. That is because $\textbf{v}_i=\textbf{c}_i+\textbf{u}_i$ includes the dynamical variable when the gas moves as a whole and in kinetic theory we desire to isolate the chaotic part of the velocity that generates thermodynamical fluxes \cite{Brush,Maxwell}.
\\

To derive the conservation equations we take Eq. (1) plus the collision conserved quantities and integrate over the velocities $\textbf{c}_i$. For the matter density we multiply by $m_i$ to get,
\begin{eqnarray}
\frac{\partial\rho_{i}}{\partial t}+\nabla\cdot(\rho_{i}\textbf{u}_{i})=0,\label{massconservation}
\end{eqnarray}
the expression for mass conservation. For the momentum balance, or Newton's second law for fluids, we multiply by $m_{i}\textbf{v}_{i}$, to get
\begin{eqnarray}
\frac{\partial}{\partial t}\left[\rho_{i}\textbf{u}_{i}\right] +\nabla\cdot\left[\tau_{i}+\rho_{i}\textbf{u}_{i} \textbf{u}_{i}\right]=0.\label{momentumconservation}
\end{eqnarray}
And for the balance for the internal energy density we multiply by $\frac{1}{2}m_{i}c_{i}^{2}$
\begin{eqnarray}
\rho_i\frac{\partial\varepsilon_{i}}{\partial t} +\nabla\cdot\textbf{J}_{q_{i}}+\rho_i\textbf{u}_i\cdot\nabla \varepsilon_i
=\int\frac{1}{2}m_if_ic^2_iJ(f_if_j)d\textbf{c}_i, \label{energyconservation}
\end{eqnarray}
where $i=a,b$. If we sum over $i$ making $T_{a}=T_{b}$, and substitute $\textbf{u}_{i}$ by $\textbf{u}$ as a local variable we recover the mixture expressions under full local equilibrium.
\\

\section{Solution Method: Chapman-Enskog Expansion}

Since the Boltzmann equation for each species Eq. (1) has the same structure as the one in which the temperatures are not different, there is no need to undertake the details about its solution, which is given in Refs. \cite{Chapman-Cowling,teoria cinetica LGCS}. The species are independent, as it can be seen from the balance equations. Thus, when we assume the validity of the functional hypothesis, namely $f_{i}\left(\textbf{r},\textbf{v}|n_{i}(\textbf{r},t), u_{i}(\textbf{r},t),e_{i}(\textbf{r},t)\right)$, we are explicitly indicating that the five independent state variables for each species are $n_i,u_i,e_i$; $i=a,b$. Next $f_i$ may be expanded in a power series of Knudsen's parameter $\epsilon$ around the local equilibrium distribution function $f_{i}^{(0)}$, namely, the well-known  Maxwellian distribution,
\begin{eqnarray}
f_i^{(0)}=n_i\left(\frac{m_i}{2\pi kT_i}\right)^{3/2} \exp\left\{-\frac{m_ic^2_i}{2kT_i}\right\}.
\label{localequilibriumdistribution}
\end{eqnarray}
These are the basis of the Chapman and Enskog method. To first order in gradients, we get that,
\begin{eqnarray}
f_{i}=f_{i}^{(0)}(1+\varphi_{i}),\label{Chapman-Enskog Expansion}
\end{eqnarray}
so that from Eqs. (\ref{BE}) and (\ref{colisionalkernel}) we have that,
\begin{eqnarray}
\frac{\partial f_{i}^{(0)}}{\partial t}+\textbf{v}_{i}\cdot\frac{\partial f_{i}^{(0)}}{\partial\textbf{r}} =f_{i}^{(0)}\left[C(\varphi_{i}) +C(\varphi_{i}+\varphi_{j})\right]\label{BE lineal}
\end{eqnarray}
where the linearized collision kernels $C(\varphi_{i})$ and $C(\varphi_{i}+\varphi_{j})$ are given by
\begin{eqnarray}
C(\varphi_{i})\equiv\int\int\int f^{(0)}_{i1}{ d}\textbf{v}_{i1}'{ d}\textbf{v}_{i}{ d}\textbf{v}_{i1} \sigma_{ii} g_{i,i1}\Delta(\varphi_{i}),
\end{eqnarray}
where as in the one component case, the subscript 1 labels one of the two identical species. Also,
\begin{eqnarray}
C(\varphi_{i}+\varphi_{j})\equiv\int\int\int f^{(0)}_{j}{ d}\textbf{v}_{i}'{ d}\textbf{v}_{j}{ d}\textbf{v}_{j}' \sigma_{ij} g_{ij}\Delta(\varphi_{i}+\varphi_{j}),
\end{eqnarray}
with
\begin{eqnarray}
\Delta(\varphi_{i})=\varphi_{i}' +\varphi_{i1}'-\varphi_{i}-\varphi_{i1}
\end{eqnarray}
and
\begin{eqnarray}
\Delta(\varphi_{i}+\varphi_{j})=\varphi_{j}' +\varphi_{i}'-\varphi_{j}-\varphi_{i}
\end{eqnarray}
\\
Due to the functional hypothesis, the left hand side of Eq. (\ref{BE lineal}) can be expressed as
\begin{eqnarray}
\frac{\partial f_{i}^{(0)}}{\partial t}=\frac{\partial f_{i}^{(0)}}{\partial n_{i}}\frac{\partial n_{i}}{\partial t}+\frac{\partial f_{i}^{(0)}}{\partial\textbf{u}_{i}}\cdot\frac{\partial\textbf{u}_{i}}{\partial t}+\frac{\partial f_{i}^{(0)}}{\partial T_{i}}\frac{\partial T_{i}}{\partial t},\label{functhyp}
\end{eqnarray}
and the same for the other component. By using Eqs. (\ref{localequilibriumdistribution}), (\ref{massconservation}), (\ref{momentumconservation}) and (\ref{energyconservation}), as well as $p_i=n_ikT_i$, the linear Boltzmann equation Eq. (\ref{BE lineal}) turns out to be
\begin{eqnarray}
\frac{m_{i}}{kT_{i}} \mathring{\textbf{c}_{i}\textbf{c}_{i}}:\nabla\textbf{u}_{i} +\left(\frac{m_{i}c_{i}^{2}}{2kT_{i}}-\frac{5}{2}\right) \textbf{c}_{i}\cdot\frac{\nabla T_{i}}{T_i} =[C(\varphi_{i})+C(\varphi_{i}+\varphi_{j})],\label{BEtosolve}
\end{eqnarray}
where $\mathring{\textbf{c}_{i}\textbf{c}_{i}} =\textbf{c}_{i}\textbf{c}_{i}-(1/3)c_{i}^{2}\mathcal{I}$. Here it is important to underline that a diffusive term which would be proportional to $\nabla n_i$ does not appear. This is a consequence of taking $p_i=n_ikT_i$ instead $p=p_a+p_b=\left(n_a+n_b\right)kT$.  Solution of Eq. (\ref{BEtosolve}) is taken as the sum of the homogeneous plus inhomogeneous part \cite{Courant,Hirschfelder-Curtiss},
\begin{eqnarray}
\varphi_{i} =\overrightarrow{\mathcal{A}_{i}}\cdot\nabla\ln T_{i}+\overrightarrow{\mathcal{A}_{j}}\cdot\nabla\ln T_{j} +\overrightarrow{\mathcal{B}_{i}}:\nabla \textbf{u}_i
+\overrightarrow{\mathcal{B}_{j}}:\nabla \textbf{u}_j
\\\nonumber
+\alpha_{1}+m_{i}\vec{\alpha}_{2}\cdot\textbf{c}_{i} +\alpha_{3}\frac{1}{2}m_{i}c_{i}^{2}.
\end{eqnarray}
Using now the subsidiary conditions \cite{Libro Plasmas,On the validity of the Onsager relations}
\begin{eqnarray}
\int f_{i}^{(0)}\varphi_{i}\left(\begin{array}{c}
m_{i}\\
m_{i}\textbf{c}_{i}\\
\frac{1}{2}m_{i}c_{i}^{2}\end{array}\right)d\textbf{c}_{i}=0,
\end{eqnarray}
in what follows, we shall omit the term  $\overrightarrow{\mathcal{B}_{i}}:\nabla \textbf{u}_i$ since by Curie's theorem \cite{Groot Mazur} does not couple with a thermal gradient. Therefore the solution to Eq. (\ref{BEtosolve}) is given by,
\begin{eqnarray}
\varphi_{i}=-\frac{\textbf{c}_{i}}{\Delta T}\mathcal{A}_{i}\nabla\Delta T-\frac{\textbf{c}_{j}}{\Delta T}\mathcal{A}_{j}\nabla\Delta T=-\left(\textbf{c}_{i}\mathcal{A}_{i}+\textbf{c}_{j} \mathcal{A}_{j}\right)\nabla\ln\Delta T,\label{Solution to BE lineal}
\end{eqnarray}
where $\Delta T=T_i-T_j$. The still unknown functions $\mathcal{A}_{i}\left(|\textbf{c}_{i}|,n_i,T_i\right)$ may be expanded in terms of the Sonine-Laguerre polynomials,
\begin{eqnarray}
\mathcal{A}_{i} =\sum_{p=1}^{\infty}a_{A}^{p}S_{3/2}^{p} \left(c_{i}^{2}\right),
\end{eqnarray}
where all properties of the coefficients $a_{A}^{p}$ have been thoroughly discussed \cite{Fisica de los Procesos Irreversibles,Plasmas Grupo II,Libro Plasmas} and there is no need to undertake further unnecessary details.
\\

We now assume that the mixture of our interest is composed by electrons and heavy ions (protons) with initial temperatures $T_e^0$ and $T_p^0$ respectively. Such an assumption allow us to consider that there is a huge difference between their masses. We consider that the ions are colder than the electrons because of the mass difference. Thus $T_e^0>T_p^0$, the superscript nought indicates their initial values.
\\

With the results here obtained we are now ready to face the problem. At some initial time $t^0$, $T_e^0>T_p^0$, there is a temperature gradient in the system which is as a whole in a non equilibrium state. Heat will flow through collisions from the mixture of electrons to that of the ions until after a certain time $\tau$ elapses, $T_e=T_p=T$ and the whole mixture is in a full local equilibrium state with a certain temperature $T$. The question is, given $T_i^0$ and $T_p^0$ what is the magnitude of $\tau$? This will be discussed in the following section.

\section{Relaxation Time of the Temperatures}

Recall that the mixture is composed by electrons and protons whose the mass ratio is about 1863, and rename the subscript as $i=e$ for electrons and $j=p$ for protons. The only way for the electrons to interchange energy with protons is when they collide. Then, due to the large difference of momentum and energy in a collision, small particles just change the sign of their velocities while large particles remain practically unaltered. This is the basis of Landau's idea, so that the basic mechanism for the thermalization are the electron-proton collisions. Whence, by using the definition of kinetic temperature,
\begin{eqnarray}
\frac{3}{2}kT_{e}=\left\langle \frac{1}{2}m_{e}c_{e}^{2}\right\rangle =\frac{1}{n_{e}}\int m_{e}c_{e}^{2}f_{e} d \textbf{c}_{e},
\end{eqnarray}
we can write
\begin{eqnarray}
\frac{ \partial}{\partial t}\Delta T=\frac{4}{3kn}\left\{ \int m_{e}c_{e}^{2}\frac{ \partial f_{e}}{ \partial t} d\textbf{c}_{e}-\int m_{p}c_{p}^{2}\frac{ \partial f_{p}}{\partial t} d \textbf{c}_{p}\right\},\label{TempDiff}
\end{eqnarray}
where $\Delta T=T_e-T_i>0$. Further we have neglected the terms corresponding to the gradients in the left hand side of the Boltzmann equation, since when using the Chapman and Enskog expansion contribute only to the second order. We also use the assumption of total ionization i.e. $n_e=n_p=(1/2)n$.
\\

Substitution of Eq. (\ref{BE}) in (\ref{TempDiff}) yields,
\begin{eqnarray}
\frac{ \partial}{ \partial t}\Delta T=&\frac{4}{3kn}\int\int\int\int\left(m_{e}c_{e}^{2} -m_{p}c_{p}^{2}\right)\left\{ f'_{e}f'_{p}-f_{e}f_{p}\right\}\\\nonumber &\times\sigma_{ep}g_{ep} { d}\textbf{c}'_{e}{ d}\textbf{c}'_{p}{ d}\textbf{c}_{e} { d} \textbf{c}_{p}.\label{Eq 28}
\end{eqnarray}
Next, we take a collision model considering electromagnetic interactions Ref. \cite{Libro Plasmas}, so as shown in this reference, we have that,
\begin{eqnarray}
&\frac{1}{n\tau}=C\equiv\int\int\sigma d\textbf{c}'_{e}d\textbf{c}'_{p}\\\nonumber &=\frac{1}{\sqrt{m_{e}}} \left(\frac{ e^{2}}{8\pi\epsilon_{0}}\right)^{2} \frac{1}{\left(kT_{e}^{0}\right)^{3/4}} \ln\left[1+16\pi^{2}\left(\frac{\epsilon_{0} \left(kT_{e}^{0}\right)}{\rm e^{2}}\right)^{3}\right]
\end{eqnarray}
that has units [length]$^3$/[time]. The initial temperature of the electrons is $T^{0}_e$ (notice that it has to be also substituted in $f^{(0)}_e$).
\\
With the help of Eqs. (\ref{Solution to BE lineal}) and (\ref{Chapman-Enskog Expansion}), after substitution in Eq. (30) one obtains that, (see appendix)
\begin{eqnarray}
\frac{\partial}{\partial t}\Delta T =-h |\nabla\Delta T|\label{differential Eq for Temp}
\end{eqnarray}
with
\begin{eqnarray}
h=-\frac{10}{k}Cnm_{e} \left[1+3\left(\frac{T_{p}^{0}}{T_{e}^{0}}\right)\right] \frac{1}{\Delta T^{0}}a_{A}^{1}\label{coef}
\end{eqnarray}
where $a_{A}^{1}$ is the coefficient that comes from the Sonine-Laguerre expansion and it is well-known \cite{Plasmas Grupo II} to be of the order of the collisional time $\tau\sim \frac{1}{nC}$. To derive last equation we had to use two approximations, the first one is that in a microscopic level, when one electron collides with a proton, the energy of both apparently remains unchanged because of the large mass difference, so $\textbf{c}_e\sim-\textbf{c}'_e$ and $\textbf{c}_p\sim\textbf{c}'_p$. The second one is that we expanded in Taylor series up to first order, the logarithm $\ln\Delta T$ around the initial temperature difference $\Delta T^{0}$.
An outline of the calculations leading to Eq. (\ref{coef}) are given in the appendix.
\\

In order to integrate Eq. (\ref{differential Eq for Temp}) for simplicity  we assume that the gradient $\nabla\Delta T$ is in the $x$ direction. The resulting equation is easily solved and the result is,
\begin{eqnarray}
\Delta T(t)=\Delta T^{0}\exp{[-ht]},\label{result}
\end{eqnarray}
note than in Eq. (\ref{result}) the $x$ dependence has disappeared due to the very particular boundary conditions. This is precisely what one expects, after a very short time the two temperatures must become the same at every space point in the mixture. The difference $\Delta T$ decays essentially as Landau's predictions.
Numerically, by using for instance $T^{0}_e=450$K, $T^{0}_p=200$K and $n\sim10^{21}$(1/m$^3$), the relaxation time is about $10^{-6}$ seconds. The relaxation time is not so sensitive to the initial difference $\Delta T^{0}$, so if we insert appropriate temperatures (we mean, non relativistic) the order of magnitude does not change. This result shows explicitly that when we use kinetic theory to first order in the gradients, the relaxation time is small, and it is far from being relevant in the hydrodynamic regime.

\section{Conclusions}

We have calculated the relaxation time using a method which is entirely different as the one used by Landau \cite{Physical Kinetics}, based upon the Fokker-Plank equation. Here we used the complete
information contained in the Boltzmann equation. In the Fokker-Plank approach the collision term in the
Boltzmann equation is substituted with an ``effective term''; the problem in such equation is that it losses the
microscopic information due to the collisions. In our method, as we have said, we imposed no assumptions
on the collision term. In fact, the collisions are studied directly to find the relaxation time. This calculation
allows us to justify form a microscopic point of view the hypothesis made in \cite{Libro Plasmas} and many other works, that assume that in a binary dilute mixture or ionized plasma the
temperatures are equal. We recall that this time is of the order of microseconds, so this process is very fast compared with those of hydrodynamic interest. To extend this calculation to a plasma, one needs to incorporate a term in Boltzmann's equation in which the Lorentz force is taken into account. Examining the solution in this case \cite{Marshall,Libro Plasmas} when weak electromagnetic fields are considered, wont affect the order of magnitude of the relaxation time.
\\

It is also very important to discuss why the diffusive force does not appear in these calculations and why no
mutual diffusion is observed. When the temperatures are equal, a diffusive force appears proportional to
the gradient of the number of particles $n$, Ref. \cite{On the validity of the Onsager relations}. Now, since $T_e\neq T_p$, from our definition of chaotic velocity $\textbf{c}_i=\textbf{v}_i+\textbf{u}_i$
we have $\langle \textbf{c}_i\rangle= 0$, implying that the mass flux of each species is zero; we do not have mutual diffusion neither for the components nor the mixture. It is really important to highlight this point, the species are independent since they have totally independent maxwellian distributions. When
the temperatures differ, an electron-proton collision produces a very small change in the energy of each one
due to the large difference of the masses. In other words, in a collision between a very heavy particle and a
very light one, the energy of them is almost unchanged. Only when the temperatures of the species become
similar, we expect to see the diffusive effects. Thus, the diffusion will appear when the temperatures of the
components are equal.

A question could be raised about the applicability of this method to study relaxation times in the case of heavy Brownian particles suspended in a fluid. Although the question is meaningful it cannot be answered by the use of a Boltzmann type equation. In fact, the heavy Brownian particles undergo erratic motion due to the effect of many collisions between the particles of the fluid and them. That is why the process is better described by stochastic equations. If one would seek to find the relaxation time that takes to the Browninan particles reach the Maxwellian distribution, one have to solve such equations. For instance, a Langevin type or a Fokker-Planck type equation.

\appendix
\section*{Appendix}
\setcounter{section}{1}

The Chapmann-Enskog expansion to first order in the gradients is
\begin{eqnarray}
f_{a}=f_{a}^{(0)}+\varphi_{a},\label{A1}
\end{eqnarray}
where
\begin{eqnarray}
\varphi_{a}=-\frac{\textbf{c}_{a}}{\Delta T}\mathcal{A}_{a}\nabla\Delta T-\frac{\textbf{c}_{b}}{\Delta T}\mathcal{A}_{b}\nabla\Delta T=-\left(\textbf{c}_{a}\mathcal{A}_{a}+\textbf{c}_{b}\mathcal{A}_{b}\right)\nabla\ln\Delta T\label{A2}
\end{eqnarray}
and $\Delta T=T_{a}-T_{b}$.

The expansion of the bilinear terms in $f_i$  appearing in equation in Eq. (30), with the use of Eq. (26) leads to the expression
\begin{eqnarray}
&f'_{a}f'_{b}-f_{a}f_{b}
=f_{a}^{(0)'}f_{b}^{(0)'} -f_{a}^{(0)}f_{b}^{(0)}\\\nonumber
&+f_{a}^{(0)'}f_{b}^{(0)'}\left(\varphi'_{a}+\varphi'_{b}+\varphi'_{a}\varphi'_{b}\right) -f_{a}^{(0)}f_{b}^{(0)}\left(\varphi_{a}+\varphi_{b}+\varphi_{a}\varphi_{b}\right)\\\nonumber
&=f_{a}^{(0)}f_{b}^{(0)}\left(\varphi'_{a}+\varphi'_{b} -\varphi_{a}-\varphi_{b}\right)\\\nonumber
&=f_{a}^{(0)}f_{b}^{(0)}\left(\begin{array}{c}
-\left(\textbf{c}'_{a}\mathcal{A}_{a}+\textbf{c}'_{b}\mathcal{A}_{b}\right)\\
-\left(\textbf{c}'_{a}\mathcal{A}_{a}+\textbf{c}'_{b}\mathcal{A}_{b}\right)\\
+\left(\textbf{c}_{a}\mathcal{A}_{a}+\textbf{c}_{b}\mathcal{A}_{b}\right)\\
+\left(\textbf{c}_{a}\mathcal{A}_{a}+\textbf{c}_{b}\mathcal{A}_{b}\right)\end{array}\right)\nabla\ln\Delta T.
\label{A3}
\end{eqnarray}
Using now the fact that protons are heavy particles (species $b$) and electrons light particles (species $a$) it follows that, in a collision $\textbf{c'}_{a}\sim-\textbf{c}_{a}$ and $\textbf{c'}_{b}\sim\textbf{c}_{b}$. Thus, the relative velocity $g_{ab}=|\textbf{c}_{a}-\textbf{c}_{b}|\sim c_{a}$. Eq. (38) together with Eq. (30) leads to,
\begin{eqnarray}
\frac{\partial}{\partial t}\Delta T =&\frac{4}{3kn}C\int\int\left(m_{a}c_{a}^{2}-m_{b}c_{b}^{2}\right) f_{a}^{(0)}f_{b}^{(0)}\\\nonumber
&\times\left(4\textbf{c}_{a}\mathcal{A}_{a}\right) c_{a}d\textbf{c}_{a}d\textbf{c}_{b}\cdot\nabla\ln\Delta T.
\label{A4}
\end{eqnarray}
Direct integration in Eq. (39), using the Sonine-Laguerre expansion, the kinetic temperature $\frac{3}{2}kT_{b}^{0}=\frac{1}{2}m_{b}\left\langle c_{b}^{2}\right\rangle$ and the fact that $n_i=\int f_{i}^{(0)}d\textbf{c}_{i}$ leads to
\begin{eqnarray}
\frac{\partial}{\partial t}\Delta T=-\frac{10}{k}Cnm_{a}\left[1+3\left(\frac{T_{b}^{0}}{T_{a}^{0}}\right)\right]a_{A}^{1}|\nabla\ln\Delta T|,
\label{A5}
\end{eqnarray}
which is Eq. (\ref{differential Eq for Temp}).

\end{doublespace}
\end{document}